\documentclass[aps,pra,preprint,groupedaddress,superscriptaddress,showpacs]{revtex4-1}
\usepackage{CJK}
\usepackage{amsmath}
\usepackage{longtable}
\usepackage{mathrsfs}
\usepackage{multirow}
\usepackage{tabularx}
\usepackage{array}
\usepackage{graphicx}
\usepackage{bm}
\begin{document}
%\begin{CJK*}{GB}{}
% Use the \preprint command to place your local institutional report
% number in the upper righthand corner of the title page in preprint mode.
% Multiple \preprint commands are allowed.
% Use the 'preprintnumbers' class option to override journal defaults
% to display numbers if necessary
%\preprint{}

%Title of paper
\title{Theoretical Calculation of the Quadratic Zeeman Shift Coefficient of the $^3P^o_0$ clock state for Strontium Optical Lattice Clock}

% repeat the \author .. \affiliation  etc. as needed
% \email, \thanks, \homepage, \altaffiliation all apply to the current
% author. Explanatory text should go in the []'s, actual e-mail
% address or url should go in the {}'s for \email and \homepage.
% Please use the appropriate macro foreach each type of information

% \affiliation command applies to all authors since the last
% \affiliation command. The \affiliation command should follow the
% other information
% \affiliation can be followed by \email, \homepage, \thanks as well.
%\author{Benquan Lu, Tingxian Zhang, Hong chang, Jiguang Li}
\author{Benquan Lu}
\affiliation{National Time Service Center, 710600 Lintong, China}
\affiliation{The University of Chinese Academy of Sciences, 100088 Beijing, China}
\author{Xiaotong Lu}
\affiliation{National Time Service Center, 710600 Lintong, China}
\affiliation{The University of Chinese Academy of Sciences, 100088 Beijing, China}
\author{Jiguang Li}
\affiliation{Institute of Applied Physics and Computational Mathematics, 100088 Beijing, China}
\author{Hong Chang}
\email{changhong@ntsc.ac.cn}
\affiliation{National Time Service Center, 710600 Lintong, China}
\affiliation{The University of Chinese Academy of Sciences, 100088 Beijing, China}
%\thanks{}
%\altaffiliation{}

%Collaboration name if desired (requires use of superscriptaddress
%option in \documentclass). \noaffiliation is required (may also be
%used with the \author command).
%\collaboration can be followed by \email, \homepage, \thanks as well.
%\collaboration{}
%\noaffiliation

\date{\today}

\begin{abstract}
The quadratic Zeeman shift coefficient of $^3P^o_0$ clock state for strontium is determined in theory and experiment. In theory, we derived the expression of the quadratic Zeeman shift of $^3P^o_0$ clock state for $^{88}$Sr and $^{87}$Sr in the weak-magnetic-field approximation. By using the multi-configuration Dirac-Hartree-Fock theory, the quadratic Zeeman shift coefficients were calculated. To determine the calculated results, the quadratic Zeeman shift coefficient of $^3P^o_0,F = 9/2,M_F = \pm9/2$ clock state was measured in our $^{87}$Sr optical lattice clock. The calculated results $C_2$ = $-23.38(5)$~MHz/T$^2$ for $^{88}$Sr and the $^3P^o_0,F = 9/2,M_F = \pm9/2$ clock state for $^{87}$Sr agree well with the other experimental and theoretical values, especially the most accurate measurement recently. As the $^1S_0,F = 9/2,M_F = \pm5/2$~-~$^3P^o_0,F = 9/2,M_F = \pm3/2$ transitions have been used as another clock transition for less sensitive to the magnetic field noise, we also calculated the quadratic Zeeman shift coefficients for the other magnetic states.
\end{abstract}

% insert suggested PACS numbers in braces on next line
% insert suggested keywords - APS authors don't need to do this

%\maketitle must follow title, authors, abstract, \pacs, and \keywords
\maketitle
%\end{CJK*}
% body of paper here - Use proper section commands
% References should be done using the \cite, \ref, and \label commands
\section{INTRODUCTION}
The precision of the $^{87}$Sr optical lattice clock has achieved the 10$^{-18}$ level~\cite{McGrew2018,Oelker2019}. In strontium optical clock, the $5s^2~^1S_0$~-~$5s5p~^3P^o_0$ clock transition which has a natural linewidth of about 1~mHz is induced by the external magnetic field or the internal magnetic field-the hyperfine interaction~\cite{Lu2017}. The magnetic or hyperfine interactions break the spatial symmetry of the electric states, and lead to a mixing of the $^3P^o_0$ clock state and other states with the same parity but different angular momenta. To obtain a high precision of the optical clock, we should carefully estimate the external field effects on the clock transition as it is susceptible to the ambient environment around atoms. The external magnetic field is one of the essential factors in evaluating the uncertainty of the clock transition frequency.

For the strontium optical lattice clock, the Zeeman shift is rooted in the interaction between the atom and the external magnetic field. This brings an inconsistent Zeeman shift between the ground and excited states in the same magnetic field, which finally causes a shift of the clock transition frequency. In the $^{88}$Sr optical lattice clock, we should estimate not only the first-order Zeeman shift, but also the second-order (quadratic) Zeeman shift. The first-order Zeeman shift, which is proportional to the magnetic field intensity, can be estimated by accurately determining the Land$\acute{e}$
$g$-factor of the clock states and the external magnetic field strength. The quadratic Zeeman shift is proportional to the quadratic Zeeman shift coefficient (QZSC) $C_2$ and the square of the magnetic field strength. In the $^{87}$Sr optical lattice clock, the first-order Zeeman shift could almost be cancelled by stabilizing the clock laser to the average frequency of $^1S_0, F = 9/2, M_F = +9/2$~-~$^3P^o_0, F = 9/2, M_F = +9/2$ and $^1S_0, F = 9/2, M_F = -9/2$~-~$^3P^o_0, F = 9/2, M_F = -9/2$ transitions. The Land$\acute{e}$ $g$-factors have been widely determined in theory and experiment~\cite{Takamoto2003,Boyd2007,Shi2015,Lu2018,Zhang2021}. To our knowledge, the values of the QZSC have been accurately measured~\cite{Ludlow2008,Westergaard2011,Falke2011,Bloom2014,Nicholson2015,Bothwell2019}, but there is no \textit{ab-initio} calculation on them until now. However, repeated determination of $C_2$ is critical for precisely determining the second-order Zeeman shift and depressing its statistical uncertainty. On the other hand, accurately calculating the value of $C_2$ can provide one of the most stringent tests of atomic structure calculations as it needs accurate atomic wavefunction.

In this work, we derived an expression of the QZSC of the $^3P^o_0$ clock state for $^{88}$Sr and $^{87}$Sr. It was found that the QZSC of the $^3P^o_0$ clock state for $^{87}$Sr depends on the magnetic quantum number $M_F$. Using the multi-configuration Dirac-Hartree-Fock (MCDHF) theory, we systematically considered the electron correlations and Breit interaction effects and calculated the QZSCs. At the same time, the QZSCs of the $^3P^o_0,F = 9/2,M_F = \pm9/2$ clock states were also measured in our $^{87}$Sr optical lattice clock. We have found that there is an excellent agreement between the calculations and the measurements. As in Refs.~\cite{Oelker2019,bowden2020improving}, the $^1S_0,F = 9/2,M_F = \pm5/2$~-~$^3P^o_0,F = 9/2,M_F = \pm3/2$ transitions were proposed as another clock transition for they are less sensitive to the magnetic field noise. Our calculations can be used for evaluating not only the second-order Zeeman shift of $^1S_0,F = 9/2,M_F = \pm9/2$~-~$^3P^o_0,F = 9/2,M_F = \pm9/2$ transitions, but also the $^1S_0,F = 9/2,M_F = \pm5/2$~-~$^3P^o_0,F = 9/2,M_F = \pm3/2$ transitions.
\section{Theoretical method}
\subsection{Zeeman shift of fine-structure levels}
In the presence of an external magnetic field $\textbf{B}$, the atomic Hamiltonian is~\cite{Li2013}
\begin{equation}\label{Eq:eq1}
H = H_{fs} + H_{m},
\end{equation}
where $H_{fs}$ is the relativistic fine-structure Hamiltonian which includes the Breit interaction and the main part quantum electrodynamical (QED) effects, and $H_m$ is the Hamiltonian for the interaction between the external magnetic field and the atom. If the magnetic field does not vary throughout the atomic system, the magnetic interaction Hamiltonian $H_m$ is expressed as~\cite{Andersson2008}
\begin{equation}\label{Eq:eq2}
H_m = (\bm{N}^{(1)} + \Delta \bm{N}^{(1)}) \cdot \textbf{\textrm{B}}.
\end{equation}
Here, the tensor operator $\textbf{\textit{N}}^{(1)}$ represents the coupling of the electrons with the field, and $\Delta\textbf{\textit{N}}^{(1)}$ is the so-called Schwinger QED correction~\cite{Andersson2008}
\begin{gather}\label{Eq:eq3&4}
\bm{N^{(1)}}=\sum_{j=1}^N\bm{n}^{(1)}(j)=\sum_{j=1}^N-i\frac{\sqrt{2}}{2\alpha}r_j\Big(\bm{\alpha}_j\bm{C}^{(1)}(j)\Big)^{(1)},\\
\Delta\bm{N^{(1)}}=\sum_{j=1}^N\Delta\bm{n}^{(1)}(j)=\sum_{j=1}^N\frac{g_s-2}{2}\beta_j\bm{\Sigma}_j,
\end{gather}
where $\bm{\alpha_j}$ and $\beta_j$ are the Dirac matrices, ${\bm \Sigma}_j$ is the relativistic spin-matrix and $g_s = 2.00232$ the $g$-factor of the electron spin corrected by QED effects.

If we choose the direction of the magnetic field as the quantization axis $z$, only the magnetic quantum number $M_J$ remains the good quantum number. The atomic states with the same magnetic quantum number and parity are mixed due to the magnetic interaction~\cite{Li2013}. In the weak-magnetic-field approximation, according to the first-order perturbation theory, the atomic state wavefunction $\left|\Gamma JM_J \right\rangle$ can be written as
\begin{equation}\label{Eq:eq5}
\left| \Gamma JM_J \right\rangle  = \sum\limits_{\Gamma' J'} {{d_{\Gamma' J'}}\left| {\Gamma' J'M_J} \right\rangle },
\end{equation}
here $\Gamma$ represents the additional quantum number for describing the electronic states uniquely, and the atomic state wavefunctions $\left| {\Gamma' J'M_J} \right\rangle$ are the eigenstates of the Hamiltonian $H_{fs}$. The expansion coefficients $d_{\Gamma' J'}$ are given by
\begin{equation}\label{Eq:eq6}
{d_{\Gamma' J'}} = \frac{\left\langle {\Gamma' J'M_J} \right| H_m \left| {\Gamma J M_J} \right\rangle}{E({\Gamma J M_J}) - E({\Gamma' J'M_J})},
\end{equation}
where $\left| {\Gamma' J' M_J} \right\rangle$ stands for the perturbing states.

The second-order perturbation to the fine-structure energy level is presented as
\begin{equation}\label{Eq:eq7}
\Delta E^{(2)} = \sum\limits_{\Gamma' J'} \frac{|\left\langle {\Gamma' J'M_J} \right| H_m \left| {\Gamma J M_J} \right\rangle|^2}{E({\Gamma J M_J}) - E({\Gamma' J'M_J})}.
\end{equation}

For the $5s5p~^3P^o_0$ clock state, the adjacent $^3P^o_1$ and $^1P^o_1$ states are considered as perturbing states, and the other states are neglected because of their fractional contribution due to large energy intervals. Therefore, the QZSC can be expressed as
\begin{equation}\label{Eq:eq8}
C_2 = \sum\limits_{s = 1,3} \frac{|\left\langle {^sP^o_1} \left| {\bm{N}^{(1)} + \Delta \bm{N}^{(1)}} \right| {^3P^o_0} \right\rangle|^2}{E({^3P^o_0}) - E({^sP^o_1})}.
\end{equation}
The magnetic interaction matrix elements are given by
\begin{equation}\label{Eq:eq9}
\begin{aligned}
&\left\langle {\Gamma J{M_J}} \right|{\bm{N}^{(1)}+\Delta \bm{N}^{(1)}}\left| {\Gamma'J-1{M_J}} \right\rangle \\
=& {\left( { - 1} \right)^{J - {M_J}}}\left( {\begin{array}{*{20}{c}}
   \begin{gathered}
  J \hfill \\
   - {M_J} \hfill \\
\end{gathered}  & \begin{gathered}
  1 \hfill \\
  0 \hfill \\
\end{gathered}  & \begin{gathered}
  J-1 \hfill \\
  {M_J} \hfill \\
\end{gathered}   \\
 \end{array} } \right)\sqrt {2J + 1} \left\langle {\Gamma J\left\| {\bm{N}^{(1)}+\Delta \bm{N}^{(1)}} \right\|\Gamma'J-1} \right\rangle \\
=& \sqrt{\frac{J^2-M^2_J}{J(2J-1)}} \left\langle {\Gamma J\left\| {\bm{N}^{(1)} + \Delta \bm{N}^{(1)}} \right\|\Gamma'J-1} \right\rangle.
\end{aligned}
\end{equation}
\subsection{Zeeman shift of hyperfine-structure levels}
The Hamiltonian of an atom with nuclear spin $I$ ($\neq$ 0) can be expressed as
\begin{equation}\label{Eq:eq10}
H = H_{fs} + H_{hfs} + H_{m},
\end{equation}
here, $H_{hfs}$ is the interaction between the electrons and the nonspherical electromagnetic multipole moments of the nucleus. The hyperfine interaction couples the total electronic angular momentum \textbf{J} and the nuclear momentum \textbf{I} to a new total angular momentum \textbf{F}, i.e. \textbf{F} = \textbf{I} + \textbf{J}. The magnetic interaction Hamiltonian can now be written as
\begin{equation}\label{Eq:eq11}
H_m = (\bm{N}^{(1)} + \Delta \bm{N}^{(1)}) \cdot \textbf{\textrm{B}} + H_m^{nuc},
\end{equation}
here the last term $H_m^{nuc}$ represents the interaction between the magnetic field and the magnetic moment of the nucleus~\cite{Andersson2008}. It is weak and can be neglected in this work.

Similarly, we choose the direction of the magnetic field as $z$ direction. According to the first-order perturbation theory, the atomic state wavefunction can be expressed as
\begin{equation}\label{Eq:eq12}
\left| {\gamma \Gamma IJF{M_F}} \right\rangle  = \sum\limits_{\Gamma' J'F'} {{d_{\Gamma' J'F'}}\left| {\gamma \Gamma' IJ'F'{M_F}} \right\rangle },
\end{equation}
here, the magnetic-field-induced mixing coefficients $d_{\Gamma' J'F'}$ are given by
\begin{equation}\label{Eq:eq13}
{d_{\Gamma' J'F'}} = \frac{\left\langle {\gamma \Gamma' IJ'F'M_F} \right| H_m \left| {\gamma \Gamma IJFM_F} \right\rangle}{E({\gamma \Gamma IJFM_F}) - E({\gamma \Gamma' IJ'F'M_F})}.
\end{equation}
Therefore, the second-order perturbation to the hyperfine energy level can be presented as
\begin{equation}\label{Eq:eq14}
\begin{gathered}
  \Delta {E^{\left( 2 \right)}} = \sum\limits_{J',F'} {\frac{{{{\left| {\left\langle { \gamma \Gamma' IJ'F'M_F } \right|{H_m}\left| { \gamma \Gamma IJFM_F } \right\rangle } \right|}^2}}}
{{{E(\gamma \Gamma IJFM_F)} - {E(\gamma \Gamma' IJ'F'M_F )}}}}  \hfill. \\
\end{gathered}
\end{equation}

For $^{87}$Sr, the $5s5p~^3P^o_0$ clock state has only one total angular quantum number $F = 9/2$, and there is no nearby hyperfine levels to mix in analyzing the second-order Zeeman shift, which is opposed to the traditional case in alkali-metal(-like) atoms. Hence, we also treat the adjacent $^3P^o_1$ and $^1P^o_1$ states as perturbing states. Therefore, the QZSC is given by
\begin{equation}\label{Eq:eq15}
C_2 = \sum\limits_{s,F'} \frac{|\left\langle {^sP^o_1,F'M_F} \left| {\bm{N}^{(1)} + \Delta \bm{N}^{(1)}} \right| {^3P^o_0,F,M_F} \right\rangle|^2}{E({^3P^o_0}) - E({^sP^o_1})}.
\end{equation}
The energy interval in the denominator is mainly from the fine-structure splitting, and the hyperfine splitting is neglected. The magnetic interaction matrix elements are given by
\begin{equation}\label{Eq:eq16}
\begin{gathered}
  \left\langle {\gamma \Gamma IJF{M_F}} \right| H_m \left| {\gamma \Gamma 'IJ'F{M_F}} \right\rangle  \hfill \\
   = {M_F}\sqrt {\frac{{2F + 1}}
{{F\left( {F + 1} \right)}}} {\left( { - 1} \right)^{I + J' + 1 + F}}\left\{ {\begin{array}{*{20}{c}}
   \begin{gathered}
  J \hfill \\
  F \hfill \\
\end{gathered}  & \begin{gathered}
  F \hfill \\
  {J'} \hfill \\
\end{gathered}  & \begin{gathered}
  I \hfill \\
  1 \hfill \\
\end{gathered}   \\
 \end{array} } \right\}\sqrt {2J + 1} \left\langle {\Gamma J\left\| {{N^{\left( 1 \right)}} + \Delta {N^{\left( 1 \right)}}} \right\|\Gamma 'J'} \right\rangle B, \hfill \\
\end{gathered}
\end{equation}
where $J' = J-1, J$.
\begin{equation}\label{Eq:eq17}
\begin{gathered}
  \left\langle {\gamma \Gamma IJF{M_F}} \right| H_m \left| {\gamma \Gamma 'IJ'F - 1{M_F}} \right\rangle  \hfill \\
   = \sqrt {\frac{{{F^2} - M_F^2}}
{F}} {\left( { - 1} \right)^{I + J' + 1 + F}}\left\{ {\begin{array}{*{20}{c}}
   \begin{gathered}
  J \hfill \\
  F - 1 \hfill \\
\end{gathered}  & \begin{gathered}
  F \hfill \\
  {J'} \hfill \\
\end{gathered}  & \begin{gathered}
  I \hfill \\
  1 \hfill \\
\end{gathered}   \\
 \end{array} } \right\}\sqrt {2J + 1} \left\langle {\Gamma J\left\| {{N^{\left( 1 \right)}} + \Delta {N^{\left( 1 \right)}}} \right\|\Gamma 'J'} \right\rangle B \hfill, \\
\end{gathered}
\end{equation}
where $J' = J-1, J, J+1$. From Eqs.~(\ref{Eq:eq16}) and (\ref{Eq:eq17}), the magnetic matrix element between states with the same $F$ values depends on $M_F$, while that with $\Delta F = 1$ depends on absolute value of $M_F$. Therefore, the value of the QZSC is $M_F$-dependent.
\subsection{MCDHF Theory}
In the framework of the MCDHF method, the atomic state function (ASF) $\Psi (\Gamma PJM_J)$ is a linear combination of configuration state functions (CSFs) $\Phi_j (\gamma_j PJM_J)$ with same parity $P$, total angular momentum $J$ and its component along $z$ direction $M_J$, that is,
\begin{equation}\label{Eq:eq18}
\Psi (\Gamma PJM_J) = \sum_{j}^{N}c_j\Phi_j (\gamma_jPJM_J).
\end{equation}
Here, $c_j$ represents the mixing coefficient corresponding to the $j^{th}$ configuration state function, and $\gamma$ stands for the other quantum numbers which can define the state uniquely. The configuration state functions $\Phi_j(\gamma_jPJM_J)$ are built from sums of products of the one-electron Dirac orbitals
\begin{equation}\label{Eq:eq19}
\phi(r,\theta ,\varphi ,\sigma) = \frac{1}{r}\binom{P(r)\chi_{\kappa m}(\theta ,\varphi ,\sigma )}{iQ(r)\chi_{-\kappa m}(\theta ,\varphi ,\sigma )},
\end{equation}
where $P(r)$ and $Q(r)$ are the radial wavefunctions. The coefficients $c_j$ and the radial functions are optimized simultaneously in the self-consistent field (SCF) procedure. Higher-order electron correlations, the Breit interaction and quantum electrodynamical (QED) corrections can be considered in the relativistic configuration interaction (RCI) computation, in which only the expansion coefficients are varied.

Our calculation was started in the Dirac-Hartree-Fock (DHF) approximation. The occupied orbitals in the reference configuration $1s^22s^22p^63s^23p^63d^{10}4s^24p^65s5p$, or called spectroscopic orbitals, were optimized and kept frozen in the following computations. The outermost $5s$ and $5p$ electrons were regarded as the valence orbitals and others as the core. In the following SCF calculations, the valence-valence (VV) and major core-valence (CV) electron correlations were taken into account. The major CV electron correlation includes those between electrons in the valence and $n \ge 3$ core shells. The virtual orbitals were added layer by layer up to $n = 11$ and $l = 4$.

Keeping all orbitals frozen, we further considered the effect of the CV correlation related to the $n \le 2$ electrons in the subsequent RCI computation. This model is labelled as CV. The core-core (CC) electron correlation in the $4s$ and $4p$ subshells, referred to as CC$_{4}$, was also captured in RCI. To control the number of CSFs, only the first five layers of virtual orbitals were used to generate the CSFs accounting for the CC correlation. Higher-order correlations among $n \ge 4$ electrons was considered by the multi-reference (MR) single (S) and double (D) excitation approach. The MR configurations are composed of \{$4s^24p^65s5p$, $4s^24p^64d5p$, $4s^24p^65s6p$, $4s^24p^65p6s$, $4s^24p^64d6p$\}. The corresponding configuration space was expanded by SD-excitation CSFs from the MR configuration set to the first five layers of virtual orbitals. Finally, the Breit interaction was evaluated based on the MR model. In practice, we employed the GRASP2K~\cite{Joensson2013} and HFSZEEMAN~\cite{Andersson2008} packages to perform the calculations.
\section{Calculation results and discussions}
In Table~\ref{tab:1}, we display the matrix elements of the magnetic interaction $\langle^{1,3}\!P^o_1\|\bm{N}^{(1)}+\Delta \bm{N}^{(1)}\|^3\!P^o_0\rangle$, the energy intervals $\Delta E$($^{3}\!P^o_0$ - $^{1,3}\!P^o_1$) and the calculated QZSC $C_2$ of the $^3P^o_0$ state for $^{88}$Sr atom with various computational models. When calculating the magnetic matrix elements, we removed those CSFs that do not interact with the reference configurations to improve the computational efficiency. However, the corrections from these removed CSFs must be considered to the energy intervals~\cite{Lijiguang2012}. From this table, it can be seen that the VV and CV electron correlations make dominant contributions to all of the physical quantities we concerned. The contributions from the
CC correlation in $n = 4$ core shell and its corresponding higher-order correlations to the QZSC are comparable with those from the VV and CV correlations. Similar to the hyperfine interaction constants of the $5s5p~^3P_{1,2}$ and $^1P_1$ states for $^{87}$Sr~\cite{PhysRevA.100.012504}, the effect of the higher-order electron correlation on $C_2$ compensates to that of the CC correlation. Therefore, both of them were included in our calculation. Moreover, it can be noticed that the Breit interaction is also significant to improve the fine-structure splitting between $^3P^o_1$ and $^3P^o_0$ states. A good agreement is found for the fine-structure splitting $\Delta E$($^{3}\!P^o_0$ - $^{3}\!P^o_1$), but the energy interval between $^1P^o_1$ and $^3P^o_1$ deviates from the NIST~\cite{NIST_ASD} value by 4\%. The deviation is attributed to the so-called $LS$-term dependence of the $5p$ valence orbital, but the contribution from the $^1P^o_1$ perturbing state is on the level of 10$^{-6}$~MHz/T$^2$. Thus, the less good energy interval between the $^1P^o_1$ and $^3P^o_0$ states does not impact on the final $C_2$ value at present accuracy.
\begingroup\squeezetable
\begin{table}
\caption{\label{tab:1}The $\Delta E(^{3}\!P^o_0$ - $^{3,1}\!P^o_1$), $\langle^{3,1}\!P^o_1\|\bm{N}^{(1)}+\Delta \bm{N}^{(1)}\|^{3}\!P^o_0\rangle$ (all of them in atomic units), and the calculated QZSC $C_2$ (in MHz/T$^2$) of the $^3P^o_0$ clock state for $^{88}$Sr atom. Numbers in square brackets stand for the power of $10$, and in parentheses for uncertainties.}
\begin{ruledtabular}
\begin{tabular}{lccccc}
Model &$\langle^3\!P^o_1\|\bm{N}^{(1)}+\Delta \bm{N}^{(1)}\|^3\!P^o_0\rangle$ & $\langle^1\!P^o_1\|\bm{N}^{(1)}+\Delta \bm{N}^{(1)}\|^3\!P^o_0\rangle$    & $^{3}\!P^o_0$ - $^{3}\!P^o_1$ & $^{3}\!P^o_0$ - $^{1}\!P^o_1$  & $C_2$(MHz/T$^2$) \\ \hline
DHF            & 0.40913        & $-$4.0914[$-$3]  & $-$8.323[$-$4] & $-$6.976[$-$2] & $-$23.90     \\
CV             & 0.40896        & $-$1.1855[$-$2]  & $-$7.750[$-$4] & $-$3.569[$-$2] & $-$25.65     \\
CC$_{4}$       & 0.40908        & $-$7.5776[$-$3]  & $-$8.368[$-$4] & $-$4.994[$-$2] & $-$23.77     \\
MR             & 0.40897        & $-$1.2016[$-$2]  & $-$8.640[$-$4] & $-$3.506[$-$2] & $-$23.01     \\
Breit          & 0.40898        & $-$1.1847[$-$2]  & $-$8.504[$-$4] & $-$3.499[$-$2] & $-$23.38(5)    \\
                                            \multicolumn{6}{c}{Theories} \\
Taichenachev {\it {et al.}}~\cite{Taichenachev2006} & & &             &                   & $-$23.3     \\
                                            \multicolumn{6}{c}{Experiments} \\
NIST~\cite{NIST_ASD}&                  &               & $-$8.512[$-$4] & $-$3.363[$-$2] &            \\
Baillard {\it {et al.}}~\cite{Baillard2007}&              &               &              &                & $-$23.3(5)  \\
\end{tabular}
\end{ruledtabular}
\end{table}
\endgroup
\section{Experimental determination of the Quadratic Zeeman Shift Coefficient}
\begin{figure}
 \includegraphics{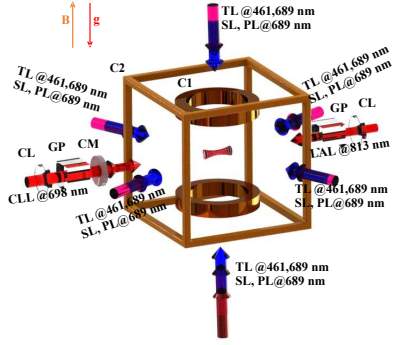}
 \caption{\label{fig:1}(color online) Schematic diagram of the optical lattice and clock transition detection of strontium optical lattice clock. C1: anti-Helmholtz coils, C2: Three-dimensional compensating coils, TL: Trapping Laser, SL: String Laser, PL: Polarizing Laser, LAL: Lattice Laser, CLL: Clock Laser, CL: Convex Lens with a focal length of 250~mm, CM: Concave Mirror, GP: Glan-Taylor Polarizer of which the polarization axis is along the direction of gravity.}
\end{figure}

The schematic diagram of clock transition detection device of $^{87}$Sr optical lattice clock is shown in Fig.~\ref{fig:1}. After two stages of cooling, the cold atoms with a population of about 10$^4$ are loaded into a horizontal one-dimensional optical lattice with a temperature of about 3~$\mu$K~\cite{Wang2018}. The 813.42~nm lattice laser is stabilized to an ultra-low expansion (ULE) cavity with a finesse of 12000. The lifetime of the atoms trapped in lattice is about 3.6~s. The clock laser is locked to another ULE cavity with a finesse of 20,0000, using the technology of Pound-Drever-Hall stabilization for repressing the frequency noise. The clock laser corresponds to the $5s^2~^1S_0$~-~$5s5p~^3P^o_0$ transition at $\lambda$ = 698~nm. The clock laser is collimated with a beam waist of 2~mm using a convex lens and overlapped with the lattice laser. The polarization of the clock laser is parallel with the lattice laser by adjusting the direction of a Glan-Taylor polarizer. Both the clock laser and the lattice laser are linearly polarized with the direction of the magnetic field and the gravity.

During obtaining clock transition spectra, the frequency of the clock laser is step changed by an Acousto-optic Modulator (AOM) to search the resonant frequency of the clock transition. The line-width of the 698~nm clock laser is about 1~Hz, and the fractional flicker noise floor of the laser instability is about 1.6$\times$10$^{-15}$. Before the measurement, the clock laser was locked to the clock transition. The interleaved instability of our clock is 4$\times$10$^{-15}$~$/\sqrt{\tau}$~\cite{Luxiaotong2019}. Considering first- and second-order terms in B, the two clock transition frequencies are given by
\begin{equation}\label{Eq:eq20}
v^{(\pm)} = v_0 \pm M_F(g_p - g_s)\frac{\mu_B}{h}B + C_2 B^2,
\end{equation}
where $v_0$ is the unperturbed resonance frequency, $g_s$ and $g_p$ are the $g$-factors for ground and excited states respectively, $\mu_B$ is the Bohr magneton, and $h$ is the Planck's constant; +($-$) refers to the $^1S_0, F = 9/2, M_F = \pm9/2$~-~$^3P^o_0, F = 9/2, M_F = \pm9/2$ transition. To measure the second-order Zeeman shift, the lock-in data must be free of the first-order Zeeman shift. This is realized by averaging systemic Zeeman sublevels and using four servos to separately lock the clock laser frequency to the corresponding transitions during the process of the self-comparison~\cite{Blatt2009}. A second integrated loop, which can obtain the slope of the drift every 68~s, is used to eliminate the frequency offset caused by the clock laser frequency~\cite{Peik2005}. Therefore, the synthesized frequency is obtained as
\begin{equation}\label{Eq:eq21}
v_{clock} = \frac{1}{2}(v^{(+)} + v^{(-)}) = v_0 + C_2 B^2.
\end{equation}
The lattice tensor shift, which may deteriorate the measurement result~\cite{Bloom2014}, is minimized by carefully aligning the polarization direction of the lattice and the clock lasers with the magnetic field quantization axis.

\begin{figure}
 \includegraphics{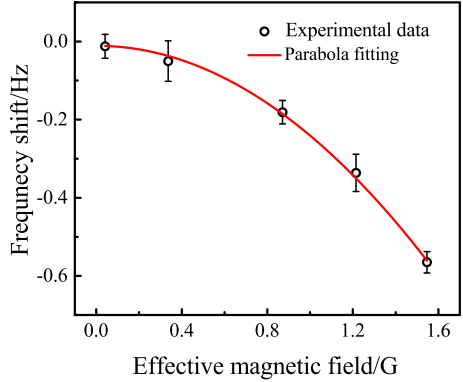}
 \caption{\label{fig:2} (color online) Relationship between second-order Zeeman shift and effective magnetic field. The error bars indicate the purely statistical 1$\sigma$ standard deviation given by the last point of the Allan deviation of self-comparison. The solid line is the parabolic fitting curve with a function of $a + b \times B^2_{eff}$.}
\end{figure}

In this measurement, we applied an adjustable bias magnetic field. The high magnetic field is denoted as $B_H$, while the low is $B_L$. The value of $B_L$ = 0.1~G remained unchanged, but $B_H$ is changed from 0.1~G to 1.55~G. To make sure that the value of the quadratic Zeeman shift vanishes when the magnetic field is zero, we defined an effective magnetic field $B_{eff} = \sqrt {B_H^2 - B_L^2}$~\cite{PhdNicholson2015}. In this way, the relationship between $B_{eff}$ and frequency shift is measured and shown in Fig.~\ref{fig:2}. In this figure, the error bars indicate the purely statistical $1\sigma$ standard deviation given by the last point of the Allan deviation of self-comparison~\cite{Yasuda2012}. The solid line is the fitting curve with a function of $a + b \times B^2_{eff}$. By parabolic fitting of the experimental data, the QZSC is obtained as $-$23.0(4)~MHz/T$^2$.

In table~\ref{tab:2}, we present the comparison of the calculated and the measured QZSC $C_2$ of the $^3P^o_0,F = 9/2,|M_F| = 9/2$ states for $^{87}$Sr. From the table, one can see that our calculated result is in good agreement with the measurements, especially the most recent measurement given by Bowden {\it {et al.}}~\cite{Bothwell2019}, the uncertainty of which is about an order magnitude smaller than the other measurements. Our measured value is also consistent with the other measurements. This determined that our computational model is reasonable.
\begin{table}
\caption{\label{tab:2} The compassion of the calculated and the measured QZSC $C_2$ (in MHz/T$^2$) of the $^3P^o_0,F = 9/2,|M_F| = 9/2$ states for $^{87}$Sr.}
\begin{ruledtabular}
\begin{tabular}{cc}
      Reference                                  &     $C_2$(MHz/T$^2$)      \\ \hline
This work(Calculation)                           &   -23.38(5)    \\
This work(Measurement)                           &   -23.0(4)    \\
Ludlow {\it {et al.}}~\cite{Ludlow2008}          &   -23.7(3)    \\
Westergaard {\it {et al.}}~\cite{Westergaard2011}&   -23.5(2)    \\
Falke {\it {et al.}}~\cite{Falke2011}            &   -23.0(3)    \\
Bloom {\it {et al.}}~\cite{Bloom2014}            &   -23.6(2)    \\
Nicholson {\it {et al.}}~\cite{Nicholson2015}    &   -23.8(8)    \\
Bothwell {\it {et al.}}~\cite{Bothwell2019}      &   -23.38(3)    \\
\end{tabular}
\end{ruledtabular}
\end{table}
\section{Conclusion}
In the weak-magnetic-field approximation, we derived an expression of the QZSC of the $^3P^o_0$ clock state for $^{88}$Sr and $^{87}$Sr. It was found that the QZSC of the $^3P^o_0$ clock state for $^{87}$Sr is $M_F$-dependent. By using the MCDHF theory, we accurately calculated the QZSCs. In our calculations, the electron correlations and Breit interaction effects were systematically considered. At the same time, the QZSCs of the $^3P^o_0,F = 9/2,M_F = \pm9/2$ clock states were also measured as $C_2$ = $-$23.0(4) MHz/T$^2$ in our $^{87}$Sr optical lattice clock. Our measurements are consistent with the other experimental values. Moreover, we have found that there is an excellent agreement between the calculations and the measurements. Our calculations can be used for evaluating not only the second-order Zeeman shift of $^1S_0,F = 9/2,M_F = \pm9/2$~-~$^3P^o_0,F = 9/2,M_F = \pm9/2$ transitions, but also the $^1S_0,F = 9/2,M_F = \pm5/2$~-~$^3P^o_0,F = 9/2,M_F = \pm3/2$ transitions. The $^1S_0,F = 9/2,M_F = \pm5/2$~-~$^3P^o_0,F = 9/2,M_F = \pm3/2$ transitions are the newly proposed clock transitions~\cite{Oelker2019,bowden2020improving} for they are less sensitive to the magnetic field noise. Our theory is also useful to predict the QZSCs for other interesting atomic systems.

\begin{acknowledgments}
This work was supported by the National Natural Science Foundation of China under Grant No. 61775220, the Strategic Priority Research Program of the Chinese Academy of Sciences under Grant No. XDB21030100, the Key Research Project of Frontier Science of the Chinese Academy of Sciences under Grant No. QYZDB-SSW-JSC004, and the West Light Foundation of the Chinese Academy of Sciences under Grant No. XAB2018B17.
\end{acknowledgments}

% Create the reference section using BibTeX:
\bibliography{mylib}

%\end{CJK*}
\end{document}